\documentclass{moriond}

\def\be{\begin{equation}}
\def\ee{\end{equation}}
\def\bea{\begin{eqnarray}}
\def\eea{\end{eqnarray}}

\def\commanderone{\texttt{Commander1}}
\def\commandertwo{\texttt{Commander2}}
\def\commanderthree{\texttt{Commander3}}
\def\commanderfour{\texttt{Commander4}}
\def\WMAP{\textit{WMAP}}
\def\Planck{\textit{Planck}}
\def\COBE{\textit{COBE}}

\def\Gaia{\textit{Gaia}}
\def\AKARI{\textrm{{AKARI}}}

\DeclareRobustCommand{\BeyondPlanck}{{\fontseries{m}\selectfont\textsc{BeyondPlanck}}}
\DeclareRobustCommand{\Commander}{%
  \texorpdfstring{\texttt{Commander}}{Commander}%
}
\DeclareRobustCommand{\commanderone}{%
  \texorpdfstring{\texttt{Commander1}}{Commander1}%
}

\usepackage[T1]{fontenc}
\usepackage{lmodern}

\usepackage{lineno}
\usepackage{wrapfig}

\newcommand{\Photo}{}

\begin{document}
\vspace*{4cm}
\title{Cosmoglobe: Mapping the Universe from the Milky Way to the Big Bang}

\author{A.I. Silva Martins, on behalf of the Cosmoglobe Collaboration}

\address{Institute of Theoretical Astrophysics, University of Oslo, Sem Sælands vei 13,\\
0371 Oslo, Norway}

\maketitle\abstracts{
The Cosmoglobe project is a global effort to jointly analyze complementary cosmological and astrophysical datasets, in order to better understand our Universe and its evolution. This paper describes the goals and motivations of the project, some of the main results and future prospects.
}

\section{Introduction}

The idea behind the Cosmoglobe project is simple: all telescopes are fundamentally looking at the same sky. By jointly analyzing many datasets at the same time, we should therefore constrain this common underlying signal, and any (instrumental or other) systematic effects that could otherwise be mistaken for cosmological signal. 

\begin{wrapfigure}{R}{0.5\linewidth}
  \vspace*{-1.7cm}
  \includegraphics[width=\linewidth]{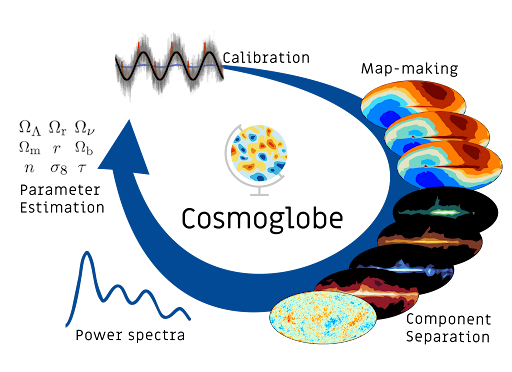}
  \caption[]{Cosmoglobe analysis loop. We start with calibrating each of the different datasets; afterwards, we make maps of the sky at each frequency that each experiment is observing; then, we separate the sky into its different components, e.g.\,the cosmic microwave background (CMB), dust, free-free, anomalous microwave emission, synchrotron, etc; additionally, we produce power spectra; and then we are able to estimate the values of cosmological parameters from these. Finally, we do it all over again, using what we have learned in the previous iteration to better calibrate the data, and so on, until we have converged to a stable solution and fully explored the viable parameter space.}\label{fig:cosmoglobe}
  \vspace*{-1cm}
\end{wrapfigure}

In order to do this, we use a Bayesian end-to-end analysis framework, called \Commander~\footnote{
``Commander is an Optimal Monte-carlo Markov chAiN Driven EstimatoR''
}~\cite{eriksen:2004,seljebotn:2019,bp03}. \Commander\ 
samples iteratively from the full joint posterior distribution of all the parameters in our model, including both the cosmological signal and the instrumental parameters. This allows us to consistently propagate uncertainties, and to marginalize over any nuisance parameters. A schematic of the Cosmoglobe analysis loop is shown in Figure~\ref{fig:cosmoglobe}. 

\section{The \Commander\ software}
\commanderone, or, at the time, simply \Commander, was originally developed as a power spectrum and component separation estimator for the high resolution maps produced by the \Planck\ mission \cite{planck2014-a12}. \commanderone\ modelled Galactic foregrounds in pixel space, and assumed all frequency channels to have the same angular resolution. While this allowed the code to run quickly, and provided great flexibility in terms of modelling spatially varying spectral energy densities (SED), it also implied that all frequency channels had to be smoothed to a common angular resolution prior to analysis. The \commanderone\ code was developed in Fortran and is still available on GitHub~\footnote{\url{https://github.com/Cosmoglobe/Commander1}}.

\commandertwo\ was the second generation of the \Commander\ framework, designed to add support for multi-resolution datasets. This code was used for the \Planck\ PR3 and PR4 data releases.

After the official end of the \Planck\ collaboration, \commandertwo\ was superseded by \commanderthree, which added support for end-to-end analysis, starting from the raw data, all the way to cosmological parameter estimation. This is the current state-of-the-art \Commander\ code, and all Cosmoglobe results to date have been produced using \commanderthree. The \commanderthree\ code is written in Fortran and is available on GitHub~\footnote{\url{https://github.com/Cosmoglobe/Commander}}, and it is still maintained and developed at the time of writing.

\commanderfour\ is the latest generation of the \Commander\ framework, and represents a significant rewrite of the legacy code. 
Whereas \commanderthree\ has a linear parallelization architecture, and only scales well within one node, \commanderfour\ is designed to run on a large number of compute nodes. This is necessary to run on modern high performance computing architectures, so that we can handle larger datasets, such as Simons Observatory and LiteBIRD. It is also more modular and flexible than its predecessors, allowing new contributors to easily analyze their own datasets or contribute new features. \commanderfour's main interface is being developed in Python, with low-level implementations in C++, and is available on GitHub~\footnote{\url{https://github.com/Cosmoglobe/Commander4}}.

\section{Global end-to-end analysis processing of Planck LFI, WMAP and COBE-DIRBE}
The Planck experiment was a huge scientific success that transformed our understanding of the early Universe, and today this data set represents a cornerstone in the current cosmological standard $\Lambda$CDM model. However, achieving those successes was highly non-trivial, and many technical challenges had to be overcome. An important result from the experiment was therefore not only the actual resulting data products, such as frequency and component maps or angular power spectra, but also many key lessons learned about how to design and analyse high-precision cosmological experiments. 

One of these lessons learned concerned the organizational structure of the analysis pipeline. In particular, the \Planck\ analysis was performed across many different institutions, each performing some part of the complete pipeline, and human interaction was usually required at many intermediate steps. For example, the Planck LFI calibration and low-level processing was performed in Trieste; mapmaking in Helsinki; and component separation in Oslo. Since the various products depended on inputs from the others, data products had to be transferred between locations repeatedly, and due to the long time lag introduced by this organization, only a handful of complete analysis cycles were completed before the end of the project.

After the official end of the Planck collaboration, the \BeyondPlanck\ project~\cite{bp01} was formed, with the goal of integrating the entire analysis loop into one code, which eventually was called \commanderthree. Once completed, this allowed performing thousands of the LFI analysis cycles without intermediate human interaction in a few months, and the first results from this analysis were published in 2021.

An important result from this work was a new estimation of the optical depth to reionization $\tau$ with full end-to-end  error propagation. 
This was also the first time a joint estimation of the full posterior distribution of all parameters, starting from data streams, and all the way to cosmological parameters, had been done, mapping all of the extremely non-trivial correlations between them, and not just their errors. 

A second major lesson learned from Planck was that the astrophysical sky is extremely rich, and it is very difficult for any one experiment to constrain all relevant parameters by itself. For instance, the \WMAP\ and Haslam data sets proved invaluable to model the low-frequency sky together with Planck. After the end of \Planck, and in parallel with the \BeyondPlanck\ project, the Cosmoglobe project was therefore born, with the ultimate goal of building one statistically coherent model of the radio, microwave and sub-millimeter sky from as many experiments as possible. The first major application of this global framework was a joint analysis of LFI and WMAP, both modelled at the time-ordered data level, and the results were published as the Cosmoglobe Data Release 1~\cite{watts2023_dr1} in 2023. This analysis finally tracked down the long-standing discrepancies between the \Planck\ and WMAP polarization maps, and showed that both experiments had struggled with estimating their own calibration accurately due to strong internal degeneracies. However, when analyzed together, these degeneracies were broken, and both experiments could, for the first time, be used jointly to constrain the polarized microwave sky.

\begin{figure}[htb]
\centerline{\includegraphics[width=\linewidth]{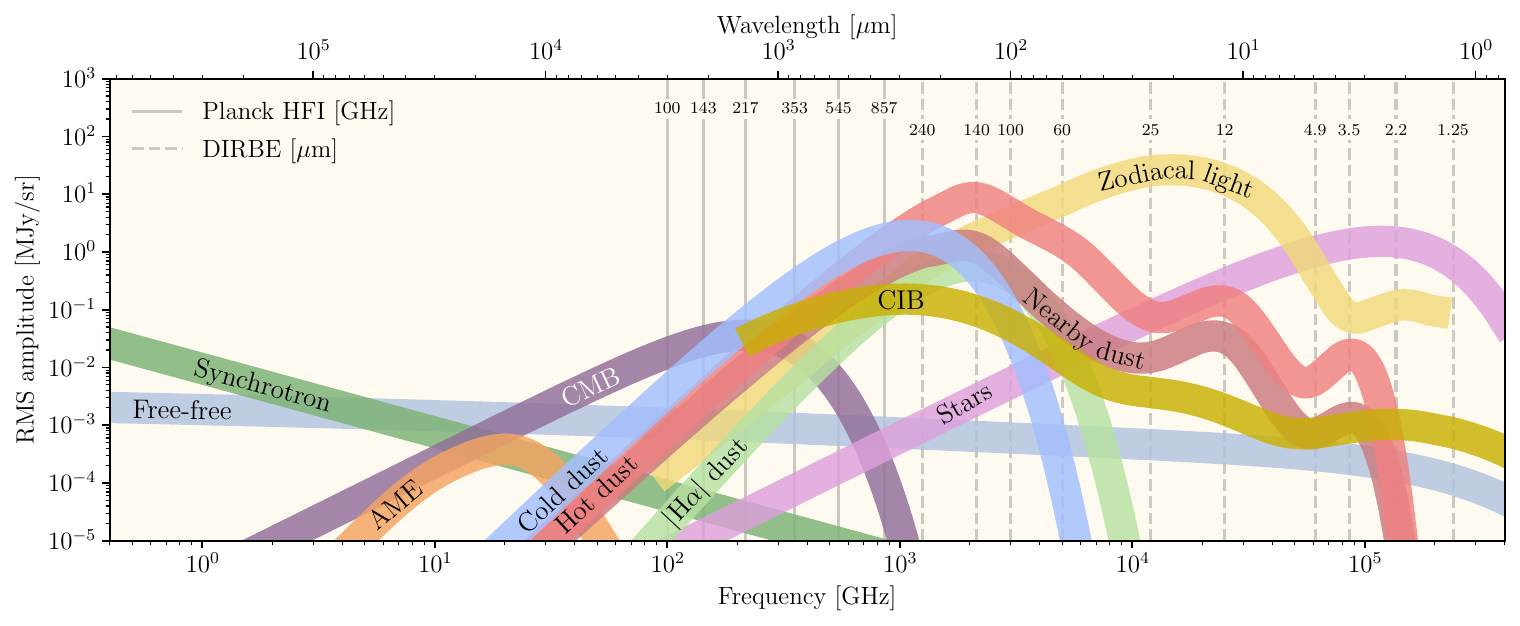}}
\caption[]{Spectral energy density overview of the sky model resulting from the Cosmoglobe DR2 analysis.}\label{fig:sed}
\end{figure}

With two successful analyses completed in the microwave frequencies, the Cosmoglobe Collaboration shifted its attention to the infrared frequencies, to better characterize the astrophysical foregrounds at higher frequencies, which, in turn, would help us get cleaner CMB maps and, thus, better estimation of the cosmological parameters that describe the Universe. In particular, the Cosmoglobe Data Release 2~\cite{CG02_01} focused on the joint analysis of the \COBE-DIRBE data, starting from the raw time-ordered data, together with information from \Gaia, \Planck\ HFI, \COBE-FIRAS, HI4PI, WHAM and Dame. This resulted in improved zodiacal-light corrected maps, and a new characterization of the thermal dust emission, together with a more complete sky model, spanning five orders of magnitude in frequency, as shown in Figure~\ref{fig:sed}.

\section{Current Status and Future Prospects}

Since its beginning, the Cosmoglobe project aimed to bring the cosmology community together, through joint analysis of datasets, and through the sharing of knowledge and, of course, data~\cite{bp05}. As such, all Cosmoglobe products can be accessed through our website~\footnote{https://www.cosmoglobe.uio.no/}.

The development of \commanderfour\ signifies that datasets that were previously inaccessible to us can now be analyzed in a reasonable amount of time. The Cosmoglobe team is then actively working on the following projects, amongs others, which we believe will benefit the community at large:
\begin{enumerate}
\item the re-analysis of \AKARI\ data, focusing on better zodiacal light modelling, and low level systematics, to characterize thermal dust emission and the cosmic infrared background at high angular resolution;
\item the re-analysis of \COBE-FIRAS data, focusing on thermometer calibration and glitch correction, to refine the measurement of the CMB temperature, and the characterization of the CMB spectral distortions;
\item the re-analysis of \Planck\ HFI data, focusing on better control of systematic effects, to exploit the full potential of the HFI polarization data;
\item the analysis of CHIPASS data, which will unlock new measurements of synchrotron;
\item the analysis of Simons Observatory data, to get high-resolution maps of the sky in the microwave, and contribute to the first measurement of the tensor-to-scalar ratio $r$;
\item the analysis of Fermi-LAT data, expanding our frequency range to the gamma-ray sky;
\item and the analysis of COMAP data, as a first step towards a joint, iterative, end-to-end analysis of line intensity mapping data.
\end{enumerate}

The Cosmoglobe team is also preparing for future experiments, such as LiteBIRD and FOSSIL, and is involved in the development of their simulations and data analysis pipelines.

\section*{Acknowledgments}
The current work has received funding from the European Union’s Horizon
  2020 research and innovation programme under grant agreement numbers
  819478 (ERC; \textsc{Cosmoglobe}), 772253 (ERC;
  \textsc{bits2cosmology}), 101165647 (ERC, \textsc{Origins}),
  101141621 (ERC, \textsc{Commander}), and 101007633 (MSCA;
  \textsc{CMBInflate}).  This article reflects the views of the
  authors only. The funding body is not responsible for any use that
  may be made of the information contained therein. This research is
  also funded by the Research Council of Norway under grant agreement
  numbers 344934 (YRT; \textsc{CosmoglobeHD}) and 351037 (FRIPRO;
  \textsc{LiteBIRD-Norway}).

\section*{References}
\bibliography{anaisabelsilvamartins}

\end{document}